\documentclass[prl,superscriptaddress,amsmath,amssymb,twocolumn,nofootinbib]{revtex4-2}

\usepackage{graphicx,bm,amsmath}
\usepackage[usenames,dvipsnames]{xcolor}
\usepackage[colorlinks,bookmarks=false,citecolor=NavyBlue,linkcolor=Red,urlcolor=blue]{hyperref}
\usepackage{amsthm}
\usepackage[bbgreekl]{mathbbol}
\usepackage{dcolumn}
\usepackage{braket}
\usepackage{comment}
\usepackage{graphicx} 
\usepackage{lipsum}
\usepackage{appendix}

\newcommand{\nsamples}{\mathcal{N} }
\newcommand{\depth}{\mathcal{D} }

\date{\today} 

\DeclareUnicodeCharacter{2212}{-}
\begin{document}

\title{Retrieving non-stabilizerness with Neural Networks}
\author{Antonio Francesco Mello}
\affiliation{International School for Advanced Studies (SISSA), 34136 Trieste, Italy}
\author{Guglielmo Lami}
\affiliation{International School for Advanced Studies (SISSA), 34136 Trieste, Italy}
\affiliation{Laboratoire de Physique Théorique et Modélisation, CY Cergy Paris Université, CNRS, F-95302 Cergy-Pontoise, France}
\author{Mario Collura}
\affiliation{International School for Advanced Studies (SISSA), 34136 Trieste, Italy}
\affiliation{INFN Sezione di Trieste, 34136 Trieste, Italy}

\begin{abstract}
Quantum computing's promise lies in its intrinsic complexity, with entanglement initially heralded as its hallmark. However, the quest for quantum advantage extends beyond entanglement, encompassing the realm of nonstabilizer (magic) states. Despite their significance, quantifying and characterizing these states pose formidable challenges.
Here, we introduce a novel approach leveraging Convolutional Neural Networks (CNNs) to classify quantum states based on their magic content. Without relying on a complete knowledge of the state, we utilize partial information acquired from measurement snapshots to train the CNN in distinguishing between stabilizer and nonstabilizer states.
Importantly, our methodology circumvents the limitations of full state tomography, offering a practical solution for real-world quantum experiments.
In addition, we unveil a theoretical connection between Stabilizer Rényi Entropies (SREs) and the expectation value of Pauli matrices for pure quantum states. Our findings pave the way for experimental applications, providing a robust and accessible tool for deciphering the intricate landscape of quantum resources.
\end{abstract}

\maketitle

\paragraph{Introduction. --} The intrinsic exponential complexity of quantum systems plays a key role within the realm of quantum computation, allowing for the possibility to overcome certain limitations of the classical counterpart in solving a variety of problems. Entanglement has been first identified as the foremost feature encoding this complexity and has henceforth been extensively studied as a crucial quantum resource~\cite{cirac2012goals, eisert2010colloquium, houck2012chip,preskill2012quantum}. 
Nevertheless, quantum advantage does not rely uniquely on entanglement. 
Indeed, Gottesman-Knill theorem shows that many quantum protocols encompassing highly entangled states can be efficiently simulated with classical resources~\cite{https://doi.org/10.48550/arxiv.quant-ph/9705052, Aaronson_2004, Gottesman_1998}. Those states, referred to as stabilizer states, are defined through the action of the Clifford group, which maps Pauli strings into Pauli strings, on the computational basis state $\ket{0\dots 0}$. 

Consequently, entanglement is not the sole resource playing a role in determining the potential of quantum computing. Nonstabilizerness, or quantum magic, has in fact been introduced as the quantity accounting for the difficulty involved in simulating a quantum non Clifford circuit on a classical device.
Several measures of nonstabilizerness, such as stabilizer nullity~\cite{beverland2020lower} and robustness of magic~\cite{heinrich2019robustness, howard2017robustness} have been proposed. These quantities, however, are generally difficult to compute even numerically since often require solving challenging optimization problems. More recently, the Stabilizer Rényi Entropies (SREs)~\cite{Leone2022} have been introduced, attracting significant attention due to the relative ease with which they can be calculated 
~\cite{Lami_2023, Lami_2024,Rattacaso_2023, tirrito2023quantifying, tarabunga2023manybody, Oliviero_2022,Haug_2023_1, Haug_2023_2} and to their suitability for experimental measurements \cite{Oliviero2022,niroula2023phase}. 
SREs possess good properties, but cannot be considered as genuine magic monotones for Rényi index $>2$ (since they are not non-increasing under projective measurements~\cite{Haug_2023_2}). Accordingly, considerable endeavours are still devoted to trying to find nicely behaved and readily evaluable quantities enabling a profound characterization of this resource. 

Quantum mechanical curse of dimensionality indeed poses a significant obstacle when dealing with such problems even when the task of quantification boils down to classification. In other words, even distinguishing between stabilizer and nonstabilizer states can turn to be complicated especially for approaches that require the full knowledge of quantum states (tomography). 

In recent years, Neural Networks (NNs) have emerged as valuable instruments in various fields of physics, ranging from condensed matter and quantum physics~\cite{PRXQuantum.2.040201} to high-energy~\cite{PhysRevResearch.4.013231}. Among the others, several efforts have been focused on their employment in phases of matter classification~\cite{Carrasquilla_2020, van_Nieuwenburg_2017,wu2023learning,Wang_2016} and entanglement detection, both in the context of supervised \cite{asif2023entanglement} and unsupervised learning~\cite{Chen_2022_unsup}. 
Another notable example concerns NN ansatz for the many-body quantum problem~\cite{Carleo_2017, Hibat_Allah_2020, PhysRevB.106.L081111,carleo2019nndyn}. Recently the problem of determining entanglement properties of a quantum state with classical artificial NN has been addressed in several works.
However, it is often assumed to possess a complete knowledge of the state~\cite{ureña2023entanglement}, thus requiring a 
full state tomography that is experimentally highly
demanding for large systems~\cite{barreiro2005tomo, chapman2016tomo,doi:10.1126/sciadv.add7131}. 
In a realistic setup, nonetheless, the state $\ket{\psi}$ of the system is known only through a limited amount of measurements outcomes. 


In this work, we introduce a novel approach to classify quantum states based on their magic content. For this purpose, we leverage a Convolutional Neural Network (CNN) to identify nonstabilizer states starting from a partial knowledge about them. In particular, we train the NN with a large set of measurement outcomes (snapshots) obtained from simple product states. Each sequence of snapshots is categorized as 'stabilizer' (0) or 'non-stabilizer' (1) based on the state's nature. Subsequently, the NN is trained to function as a binary classifier following the supervised learning approach, and its performance is assessed across diverse scenarios. 
Notably, the model's predictions are found to be robust under Clifford evolution, proving reliable even in presence of entanglement. This approach hence provides an accessible and versatile black-box tool being remarkably suitable for experimental applications.
As a side effect we also establish a theoretical direct connection between SREs and the expectation value of a string of Pauli matrices for a pure quantum state. \\

\paragraph{Preliminaries. --}
Let us consider a $N$-qubits system. The Pauli group $\mathcal{P}_N$ is defined as the set of all $N$ Pauli strings taken with proper phase factors: $\mathcal{P}_N = \lbrace \pm \left( i \right) \lbrace \sigma^0, \sigma^1, \sigma^2, \sigma^3 \rbrace^{\otimes N} \rbrace$, where we denote by $\lbrace \sigma^{\alpha} \rbrace\ \left( \alpha = 0,1,2,3 \right)$ the set of Pauli matrices ($\sigma^0 = \mathbb{1}$). The $N$-qubits Clifford group is the normalizer of $\mathcal{P}_N$, meaning that $\mathcal{C}_N = \lbrace V \in U_{2^N} | V\mathcal{P}_NV^\dagger = \mathcal{P}_N\rbrace$ \cite{Nielsen_Chuang_2010}. $\mathcal{C}_N$ is generated by the set of unitaries $\lbrace CNOT, H, S\rbrace$ \cite{Gottesman_1998}. The latter, nonetheless, does not constitute a universal set for quantum computation. Indeed, in order to achieve universality this set needs to be complemented through the addition of the non Clifford $T = \sqrt{S}$ gate \cite{Nielsen_Chuang_2010}. Given a subgroup $G$ of $\mathcal{P}_N$, a stabilizer subspace can be defined as $S_G = \lbrace \ket{\psi} \text{ s.t. } M \ket{\psi} = \ket{\psi} \forall M \in G \rbrace$. If $S_G$ is one dimensional, it identifies a single state which is named stabilizer state. As mentioned, starting from the reference basis state $\ket{0 \dots 0}$ which is a stabilizer, Clifford circuits ensure classically simulable states \cite{gottesman1998heisenberg}. A state $\ket{\psi}$ is indeed a stabilizer if and only if $\ket{\psi} = C |0\dots 0 \rangle$ with $C \in \mathcal{C}_N$. 

In the quest for experimentally viable techniques to design magic witnessess relying on readily available state specifics, measurements stand out as the most natural operations to be performed on quantum systems and CNNs as promising candidates to handle the classification task.


Notably SREs, which are given by $M_\alpha(\ket{\psi}) = \frac{1}{1-\alpha}\log{\left(\sum_{\pmb{\sigma}\in \mathcal{P}_N}\frac{1}{2^N}\bra{\psi}\pmb{\sigma}\ket{\psi}^{2\alpha}\right)}$~\cite{Leone2022}, can be seen as the Rényi Entropies for the probability distribution on the set of Pauli strings given by
\begin{align}\label{paulipdf}
    \Pi_{\rho} \left( \pmb{\sigma} \right) = \frac{1}{2^N}\textnormal{Tr}\left[ \rho \,\pmb{\sigma}\right]^2
\end{align}
with $\pmb{\sigma}$ being a string of Pauli matrices and $\rho = |\psi\rangle\langle\psi |$ a pure state.

As a consequence, in order to enhance the deployed neural network with relevant features and achieve precise state classification, we employ two alternative methods for assessing information from the state:
\begin{enumerate}
    \item 
    by sampling the many-body wave function according to its probability in the computational basis;
    \item
    by sampling the state over the Pauli string configurations according to $\Pi_{\rho} \left( \pmb{\sigma} \right)$.
\end{enumerate}

\begin{figure}[!t]
\centering
\includegraphics[scale=0.47]{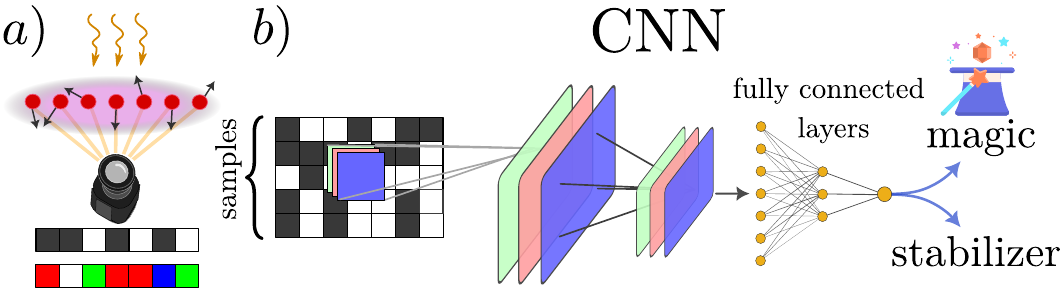}
\caption{$a)$ We consider the case of a pure quantum state $\ket{\psi}$ on a $N$-qubits system, undergoing either projective measurements of the $z$ component of the spin $S^z = \frac{1}{2}\sigma^3$ or in the Pauli basis $\lbrace \sigma^{\alpha}\rbrace$. $b)$ A Convolutional Neural Network (CNN) is employed to process the results of projective measurements. The convolutional layer outputs are then flattened and passed as input to a feedforward neural network with fully connected layers, ultimately generating the prediction outcome.\label{fig:sketch}
}
\end{figure}

\paragraph{Network technicalities. --}

Similarly to Ref.~\cite{bohrdt2021analyzing}, we use a CNN consisting of three convolutional layers with $16, 32, 64$ learnable filters respectively and ReLU activations. Input shape and specifics vary according to the chosen measurement basis (see Methods).
The output volume is flattened and put through a sigmoid classifier. Its outcomes, as a standard practice, are interpreted as probabilities for the states to be stabilizer or not. The CNN is trained with Adam optimizer~\cite{kingma2014adam} and 
binary crossentropy is chosen as loss function.
Overfitting is prevented by implementing L2 and dropout regularization schemes. Hyperparameters, including the number of epochs and the batch size, are properly tuned to speed learning up. \\

\paragraph{Methods 1 - computational basis sampling. --}
Let $\ket{\psi}$ be the normalized state for a $N$-qubits system. The total Hilbert space is $\mathcal{H} = \bigotimes_{j=1}^N \mathcal{H}_j$, where $\mathcal{H}_j$ is the local Hilbert space spanned by the eigenstates $\lbrace \ket{0}, \ket{1}\rbrace$ of the Pauli $\sigma^3=\textnormal{diag}\left( 1,-1\right)$.
An orthonormal basis for $\mathcal{H}$ is $\lbrace \ket{s_1 \dots s_N} \rbrace = \lbrace \ket{\pmb{s}} \rbrace$ with $s_j \in \lbrace 0,1\rbrace$, therefore the state of the system can be written as $\ket{\psi} = \sum_{\pmb{s}} \psi_{\pmb{s}} \ket{\pmb{s}}$. Our first investigation is based on sampling $\nsamples$ configurations from the state $\ket{\psi}$ according to the probability distribution given by $|\psi_{\pmb{s}}|^2$. The snapshots of the analyzed system correspond to real measurements in the computational basis on each qubit as depicted in Fig.~\ref{fig:sketch} $a)$. After collecting a batch of samples, they can be arranged in a $\nsamples \times N$ binary matrix, with different rows corresponding to different snapshots, and associated to the respective source state. 
The training dataset consists of random product states and random tensor products of single qubit stabilizer states, thus enhancing the feasible applicability of the proposed protocol. Snapshots originating from random product states are labelled as magic states, whereas the others are labelled as stabilizer states. 
As shown in Fig.~\ref{fig:sketch} $b)$, the core of the approach lies in using the sampled one-channel images to feed a CNN.

After completing the training process, a central test of the efficient learning of the distinctive features of nonstabilizerness is conducted by evaluating the model's performance on states undergoing random Clifford evolutions of increasing depth. Since quantum magic is Clifford invariant, a well trained model is expected to produce robust predictions - at least for shallow circuits - on snapshots of states subject to Clifford dynamics.

\begin{figure}[!t]
\centering
\includegraphics[scale=0.45]{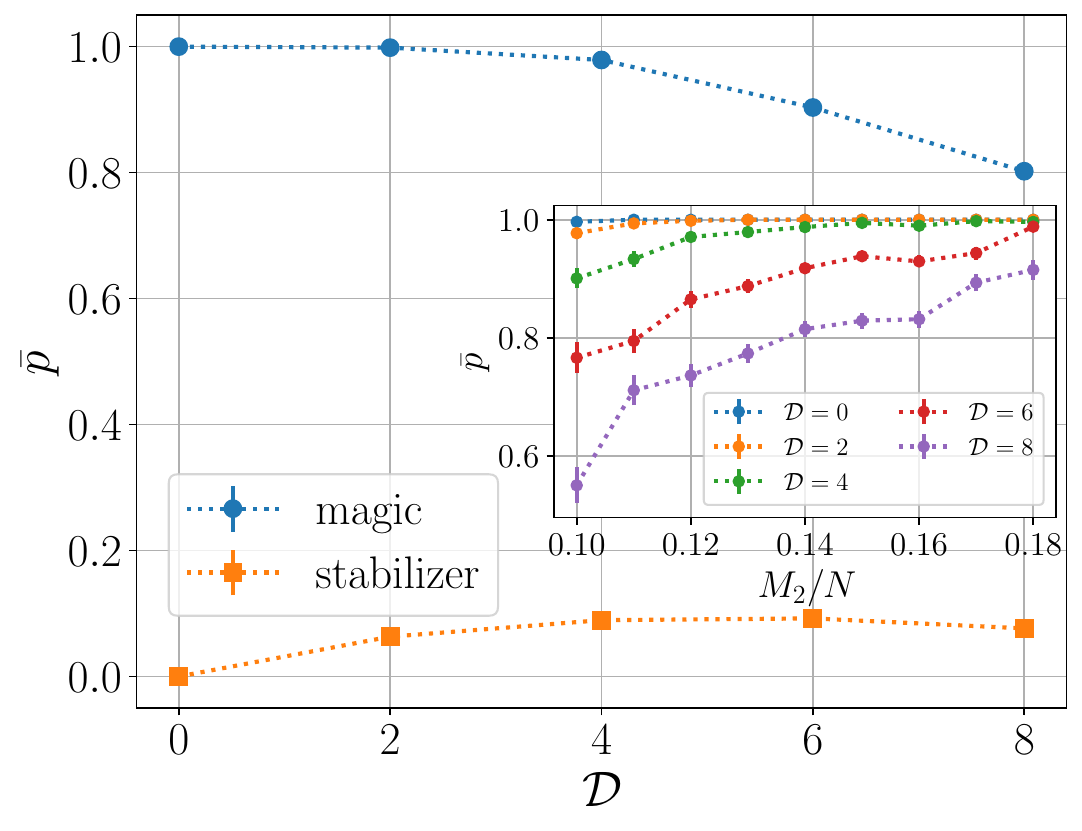}
\caption{CNN's average predictions on $m=5000$ magic and stabilizer product states undergoing random Clifford evolution of increasing depth $\depth$ for a $N=18$ qubits system sampled in the computational basis. Error bars are evaluated from the standard deviation of the average prediction $\bar{p}$. \\ Inset: average predictions as function of the density of magic of the input states. Higher values of $M_2/N$ favour the model in the classification task, specifically for larger depths $\depth$.\label{fig:clifford_ev}
}
\end{figure}

We apply such protocol to a system of size $N=18$ qubits. Specifically, we build a training set consisting of $10000$ product states, equally divided between stabilizer and nonstabilizer. After sampling $\nsamples=1000$ configurations for each state, we arrange them in $\nsamples\times N$ B/W images and feed the CNN. Learning is completed successfully, achieving satisfactory accuracy. Fig.~\ref{fig:clifford_ev} shows the performance of the CNN on samples deriving from product states undergoing random Clifford circuits of increasing depth $\depth$ (with $\depth=0$ corresponding to product states different from the ones seen during training). The number of samples and of evolving states, as well as the equal partitioning between categories, are left unchanged. 

The model is found to correctly generalize to unseen snapshots sampled from new product states ($\depth =0$). A more remarkable result, however, lies in the robustness of the network's classification under random Clifford evolution. The average predictions in fact prove reasonably reliable also for $\depth >0$, allowing to assess the quantum resource content even of arbitrarily entangled states.
Focusing on second order SREs, the inset in Fig.~\ref{fig:clifford_ev} shows that higher values of $M_2$ in the sampled states favour the CNN in correctly classifying them. 

Computational basis snapshots can also be analyzed without necessarily relying on a neural network, as will be discussed in the following paragraph. \\

\paragraph{Pauli averages as nonstabilizerness witness.--}
For any stabilizer $\ket{\psi}$ and $\pmb{\sigma} \in \mathcal{P}_N$ it holds $\bra{\psi} \pmb{\sigma} \ket{\psi} \in \lbrace 0,\pm 1\rbrace$ \cite{garcia2014simulation}. If snapshots in the computational basis are provided, one can estimate the expectation values of $\sigma^3_j$ through local sample averages. A na\"ive classifier can operate by verifying whether these estimators for each $j$ are compatible with $0,\pm 1$ within the statistical uncertainty (resulting in the classification 'stabilizer'), or if they do not (resulting in the classification 'non-stabilizer'). Yet, there are states which are misclassified by this tool, e.g. those of the form $\ket{\psi} = \otimes^N_{j=1}\frac{1}{\sqrt{2}}\left[|0\rangle + e^{i\phi}|1\rangle \right]$. One approach is to apply a shallow Clifford circuit to the state, and to repeat the classification with snapshots from the new state. This effectively allows access to $\bra{\psi} \pmb{\sigma} \ket{\psi}$, where $\pmb{\sigma} = U^{\dag} \sigma^3_j U$, and $U \in \mathcal{C}_N$. 
This classifier, operating through the evaluation of $U^\dagger \sigma^3_j U$, demonstrates remarkable performance by achieving nearly perfect classification accuracy, surpassing traditional neural networks.
Moreover, it's possible to establish a direct connection between the expectation values of Pauli strings $\pmb{\sigma}$, and the concept of non-stabilizerness.
In fact, for any $\pmb{\sigma} \in \tilde{\mathcal{P}}_N = \mathcal{P}_N/U(1) $ with $\pmb{\sigma} \neq \mathbb{1}^{\otimes N}$, 
one gets (see Supplemental Material for details)
\begin{align}\label{clifford_eq}
    \mathbb{E}_{U\in \mathcal{C}_N} \left[ \left( \bra{\psi} U^\dagger \pmb{\sigma} U \ket{\psi}\right)^4\right] =  \frac{1}{d+1} - \frac{d}{d^2-1} \, M_{\text{lin}} \left( \ket{\psi}\right),
\end{align}
where $d=2^N$ and $M_{\text{lin}} \left( \ket{\psi} \right) = 1-d\lVert \Pi_\rho\rVert^2$ is the stabilizer linear entropy linked to $M_2$ through $M_2 = -\log{\left( 1-M_{\text{lin}}\right)}$,
and we defined $\Pi_{\rho}$ as the vector of probabilities in Eq.~\eqref{paulipdf}.
Nevertheless, conducting experimental assessments of $\bra{\psi} U^\dagger \pmb{\sigma} U \ket{\psi}$ remains challenging, whether for classification purposes or for explicitly evaluating Eq.~\eqref{clifford_eq}. Clifford unitaries cannot in fact be applied to computational basis snapshots and, although in principle Pauli strings transform easily under elements of $\mathcal{C}_N$, computing their expectation value would in general result in a costly procedure requiring the full state knowledge. \\

\paragraph{Methods 2 - Pauli sampling. --}
As seen, Pauli strings allow to directly handle the action of Clifford unitaries. Hence the basis of Pauli matrices provides an alternative approach to gain valuable insight regarding the magic content of quantum states. 
In light of this, we expand our strategy to encompass a different setup, in which $\nsamples$ Pauli snapshots from the state $\rho = \ket{\psi}\bra{\psi}$ are sampled according to $\Pi_{\rho} \left(\pmb{\sigma} \right)$. As shown in the lower part of Fig.~\ref{fig:sketch} a), the latter can in turn be regarded as $4-$channels images, with each colour corresponding to a different Pauli matrix, that can be used to feed our CNN. An important advantage of this strategy is that it allows us to partially enforce Clifford invariance into the learning process. The training dataset is indeed built starting from random product states and random tensor products of single qubit stabilizer which are sampled in the Pauli basis and evolved through $l$ Clifford layers. Contrarily to the computational basis case, within this approach Clifford evolution can be effectively performed on Pauli snapshots using \cite{stim}, leveraging the tableau formalism. 

The CNN receives as input an array of shape $ \nsamples \times l \times N$ whose slices refer to different steps of the random Clifford evolution. In other words, each instance presented to the architecture intrinsically provides information about a portion of the Clifford orbit associated to the initial product state. The employment of $3-$dimensional inputs allows to extend further the testing process, accessing deeply evolved states. Additionally, product states based learning along with handily implemented Clifford evolution preserve the straightforward realization of the procedure.

In the context of this scheme, we consider $10000$ stabilizer and magic product states for $N=8,12,14$ qubits systems and sample $\nsamples = 1000$ Pauli strings for each of them. In order to construct our training set, we let these $N \times \nsamples$ images evolve under $l  = 5$ Clifford layers, thus resulting in the desired $\nsamples \times l\times N$ inputs. In complete analogy with the previous testing procedure, we evaluate the model's performance on samples collected from $10000$ product states evolved through random Clifford circuits of increasing depth $\depth$ and on which $l=5$ random Clifford layers are applied.
 Remarkably, the trained model is $l-$independent since a global pooling layer is added before the final sigmoid classifier ~\cite{cantori_scalableCNN}. In essence, once a set of snapshot has been sampled for a given state, it can undergo an arbitrarily deep Clifford circuit before being passed to the CNN.

 \begin{figure}[!t]
\centering
\includegraphics[scale=0.45]{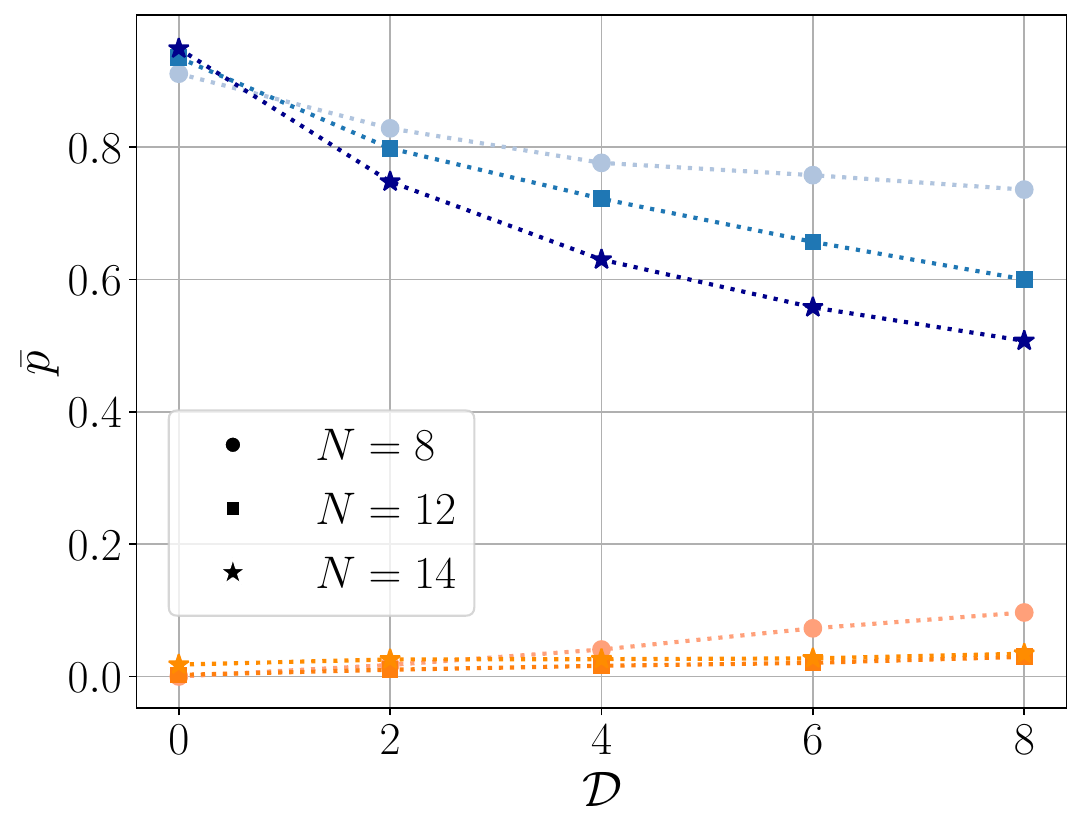}
\caption{CNN's average predictions on $m=5000$ stabilizer and magic states undergoing random Clifford evolution of increasing depth $\depth$ for $N=8,12,14$ qubits systems sampled in the Pauli basis. Input volumes are obtained through $l = 5$ Clifford layers evolution , but the same NN can be employed for deeper evolutions. Blue and orange lines respectively refer to magic and stabilizer states. Error bars are evaluated from the standard deviation of the average prediction $\bar{p}$.}
\label{fig:paulisampling_Nall}
\end{figure}
Fig.~\ref{fig:paulisampling_Nall} shows that even for large systems, where in principle a proper description demands substantial information, the CNN exhibits satisfactory generalization capabilities. Besides the noteworthy predictions for $\depth=0$ (test set of unseen product states), it also yields appropriate outcomes on deeply Clifford evolved and entangled states, thereby strengthening the validity of such protocol.

\paragraph{Conclusions and outlook. --}
We introduced a novel approach to deal with nonstabilizer states classification. In our protocol, readily obtainable snapshots of random quantum states are used to train a CNN. We showed that snapshots in the computational basis are applicable in the context of both training a neural network and devising a na\"ive magic witness based on the evaluation of $\bra{\psi} \sigma^3_j\ket{\psi}$. Although the latter is hindered by the requirement of complete wavefunction knowledge for applying Clifford layers, further investigation has been conducted. Indeed, an analytical relation between expectation values of a Pauli string and SREs for a pure quantum state has been derived. Additionally, the challenge of applying Clifford unitaries has been circumvented by resorting to sampling the state in the Pauli basis.
We have hence shown that a CNN can be trained with a limited amount of information on simple states and employed as a black-box tool to single out magic states across various scenarios. Starting from product states, the CNN has in fact been able to provide reliable predictions also on entangled states. Furthermore, the choice of Pauli basis allowed to explore sufficiently large portions of the Clifford orbit.

Our work sets the stage for novel and cost effective studies on quantum resources, bridging the gap between numerical computations and experimental applications. 
The protocol may be used to efficiently gain a deeper insight on quantum states in different settings. Moreover, the classification task may be extended to a quantification one.


\paragraph{Acknowledgments. --}
We are particularly grateful to M. Dalmonte, E.Tirrito and P. Tarabunga for inspiring discussions. This work was supported by the PNRR MUR project PE0000023-NQSTI, and by the PRIN 2022 (2022R35ZBF) - PE2 - ``ManyQLowD''. G.L. were partially founded by ANR-22-CPJ1-0021-01.\\

\bibliography{main}

\onecolumngrid

\newpage

\begin{appendices}
\section{SUPPLEMENTAL MATERIAL}
We here provide a proof of equation $\eqref{clifford_eq}$. 
We rewrite the expectation value as follows
\begin{align}
\mathbb{E}_{U \in \mathcal{C}_N} \left[ \left( \bra{\psi} U^{\dagger} \pmb{\sigma} U \ket{\psi} \right)^4\right] &= \mathbb{E}_{U \in \mathcal{C}_N} \left[ tr\left( \ket{\psi}\bra{\psi}^{\otimes 4} U^{\dagger^{\otimes 4}} \pmb{\sigma}^{\otimes 4} U^{\otimes 4}\right)\right] \nonumber \\ 
&= \mathbb{E}_{U \in \mathcal{C}_N} \left[ tr\left( \pmb{\sigma}^{\otimes 4} U^{\otimes 4} \ket{\psi}\bra{\psi}^{\otimes 4} U^{\dagger^{\otimes4}}\right)\right] \nonumber \\ 
&= tr\left( \pmb{\sigma}^{\otimes 4} \mathbb{E}_{U \in\mathcal{C_N}}\left[ U^{\otimes 4} \ket{\psi}\bra{\psi}^{\otimes 4} U^{\dagger^{\otimes 4}}\right]\right).\label{eq:clifford_average}
\end{align}
It is now possible to use the fact that \cite{tirrito2023quantifying}
\begin{align*}
\mathbb{E}_{U \in \mathcal{C}_N} \left[ \left( U \ket{\psi}\bra{\psi}U^{\dagger}\right)^{\otimes 4}\right] = \alpha Q P^{(4)}_{symm} + \beta P^{(4)}_{symm}
\end{align*}
where $\displaystyle Q = \frac{1}{4^N} \sum_{\pmb{\sigma} \in \mathcal{P}_N} \pmb{\sigma}^{\otimes 4}$ and $\displaystyle P^{(4)}_{symm} = \frac{1}{4!} \sum_{\pi \in S_4} T_{\pi}$, with $P^{(4)}_{symm}$ being the projector onto the symmetric subspace of the symmetric group $S_4$ and $T_\pi$ unitary representations of $\pi \in S_4$. The factors $\alpha$ and $\beta$ are
\begin{align*}
    \alpha &= \frac{6d(d+3) ||\Pi_{\rho}||^2-24}{(d^2-1)(d+2)(d+4)},\\
    \beta &= \frac{24(1-||\Pi_{\rho}||^2)}{(d^2-1)(d+2)(d+4)} \, .
\end{align*}
with $d=2^N$. Now we have to evaluate the following two traces
\begin{enumerate}
    \item $tr\left( \pmb{\sigma}^{\otimes 4} P^{(4)}_{symm} \right)$
    \item $tr\left( \pmb{\sigma}^{\otimes 4}Q P^{(4)}_{symm}\right)$.
\end{enumerate}

Let us start from the term $1$. 
By direct evaluation of the $4!=24$ contributions of the $T_\pi, \pi \in S_4$, it is easy to find
\begin{align}\label{eq:term1_1}
\begin{split}
     &tr\left( \pmb{\sigma}^{\otimes 4} P^{(4)}_{symm}\right) = \\
    &= \frac{1}{24} \big( \left[tr(\pmb{\sigma}) \right]^4+6 tr(\pmb{\sigma}^2)\left[tr(\pmb{\sigma})\right]^2+8tr(\pmb{\sigma})tr(\pmb{\sigma}^3)+3tr(\pmb{\sigma}^2)tr(\pmb{\sigma}^2)+6tr(\pmb{\sigma}^4) \big)  
\end{split}
\end{align}
and, if one assume $\pmb{\sigma} \neq \mathbb{1}^{\otimes N}$, 
\begin{align*}
tr\left(\pmb{\sigma}^{\otimes 4} P^{(4)}_{symm}\right) = \frac{1}{8}\left(d^2+2d \right) \, .
\end{align*}
For the term $2$, we have
\begin{align}\label{eq:term2_1}
    tr\left( \pmb{\sigma}^{\otimes 4}QP^{(4)}_{symm}\right) 
    =  \frac{1}{d^2} \sum_{\pmb{\sigma}' \in \mathcal{P}_N} tr \left(  \pmb{\sigma}^{\otimes 4} \pmb{\sigma}'^{\otimes 4} P^{(4)}_{symm}\right) \, .
\end{align}
The summation in last line of Eq.\ref{eq:term2_1} can be performed by splitting the two contribution $\pmb{\sigma} =  \pmb{\sigma}'$ and $\pmb{\sigma} \neq \pmb{\sigma}'$. The first gives $tr \left( P^{(4)}_{symm} \right)$, which is (using Eq.\ref{eq:term1_1}) \cite{mele2023introduction}:
\begin{equation*}
   tr \left( P^{(4)}_{symm} \right) = \frac{1}{24} \big( d^4+6 d^3+11d^2+6 d\big) = \binom{d+3}{4}. 
\end{equation*}
For the second contribution, as $\pmb{\sigma} \pmb{\sigma}' \in \mathcal{P}_N$ for the group closure property, we can use Eq.\ref{eq:term1_1}. 
Finally we get 
\begin{align}\label{eq:term2_2}
    tr\left( \pmb{\sigma}^{\otimes 4}QP^{(4)}_{symm}\right) &= \frac{1}{d^2} \bigg( (d^2 - 1) \frac{1}{8}(d^2+2d) + \binom{d+3}{4} \bigg) = \frac{1}{6} (1 + d) (2 + d) \, .
\end{align}
Combining all together, we obtain 
\begin{equation}
   \mathbb{E}_{U\in \mathcal{C}_N} \left[ \left( \bra{\psi} U^\dagger \pmb{\sigma} U \ket{\psi}\right)^4\right] =  \frac{1}{d+1} - \frac{d}{d^2-1} \, M_{\text{lin}} \left( \ket{\psi}\right). 
\end{equation}

\begin{figure}[!htb]
\centering
\includegraphics[scale=0.45]{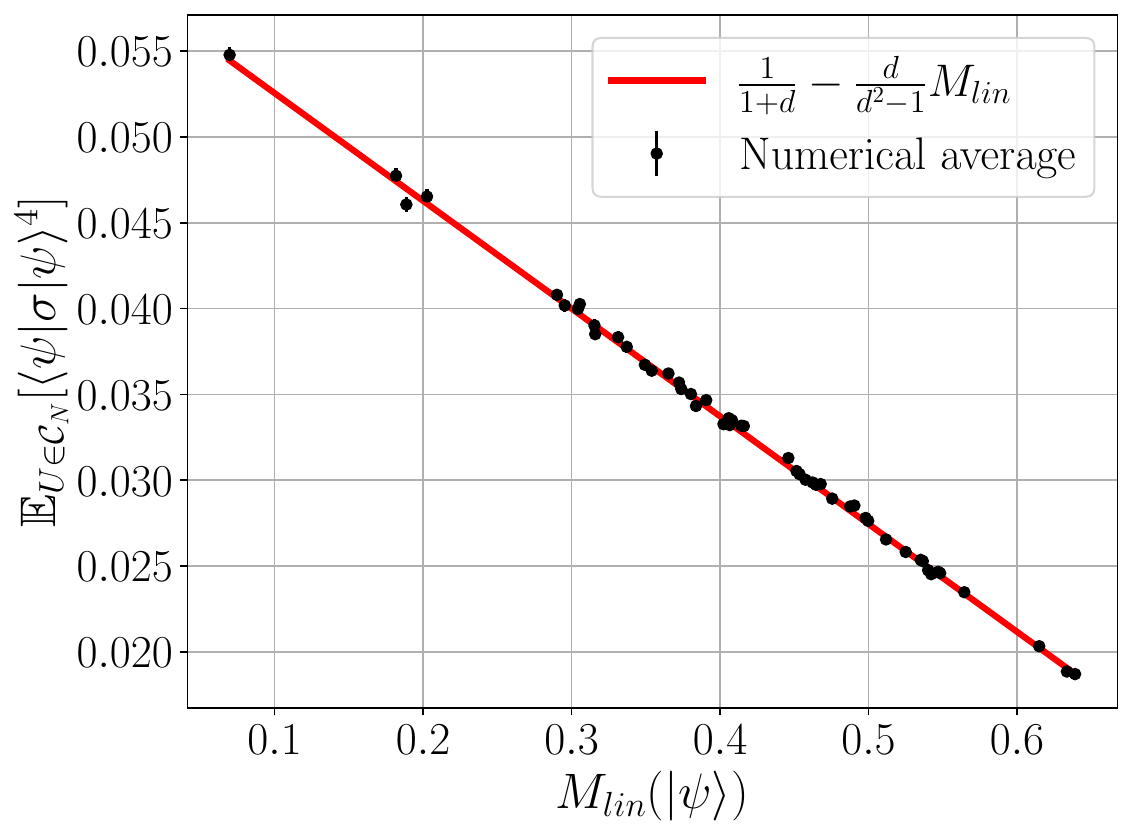}
\caption{Clifford average $\mathbb{E}_{U\in \mathcal{C}_N} \left[ \left( \bra{\psi} U^\dagger \pmb{\sigma} U \ket{\psi}\right)^4\right]$ for $50$ random product states starting from a random Pauli string $\pmb{\sigma}$ as a function of their stabilizer linear entropy $M_{\text{lin}}$. Error bars are evaluated from the standard deviation of the average.}\label{fig:clifford_averages}
\end{figure}

Figure \ref{fig:clifford_averages} shows a comparison between the numerical evaluation of the Clifford average and the one predicted by \eqref{clifford_eq} for a $N=4$ qubits system. The numerical evaluation of the average is performed for $50$ random product states, harnessing the direct application of Clifford unitaries on Pauli strings.

\end{appendices}
\end{document}